# Electromagnetic interaction between a metallic nanoparticle and surface in tunnelling proximity – modelling and experiment

J. Mitra, Lei Feng, Michael G. Boyle, and P. Dawson
Nanostructured Media Research Division, School of Mathematics and Physics,
Queen's University Belfast BT7 1NN, United Kingdom

Email: j.mitra@qub.ac.uk

We simulate the localized surface plasmon resonances of an Au nanoparticle within tunneling proximity of a Au substrate and demonstrate that the modes may be identified with those responsible for light emission from a scanning tunneling microscope. Relative to the modes of an isolated nanoparticle these modes show significant red-shifting, extending further into the infrared with increasing radius, primarily due to a proximity-induced lowering of the effective bulk plasmon frequency. Spatial mapping of the field enhancement factor shows an oscillatory variation of the field, absent in the case of a dielectric substrate; also the degree of localization of the modes, and thus the resolution achievable electromagnetically, is shown to depend primarily on the nanoparticle radius with only a weak dependence on wavelength.

## 1. Introduction

The electromagnetic interaction of light with noble metal nanostructures is of considerable current interest both from a technological as well as a fundamental physics point of view. The strong electric field enhancement associated with localized surface plasmon (LSP) excitations on such nanostructures plays a key role in various fields such as surface enhanced Raman spectroscopy (SERS) [1], biological sensing [2,[3] and targeted drug delivery [4]. As observed in the case of SERS and tip enhanced Raman spectroscopy (TERS), if the energy of the LSP modes can be pre-tuned to the absorption band of the sample under study, it leads to dramatic enhancement of the signal. Other novel applications of LSPs pertain to near-field focusing for sub-wavelength imaging [5] and plasmonic nanolithography [6,[7]. Thus it is vital to understand the interaction of electromagnetic fields with nanostructures, leading to LSP excitation. The key characteristics of the LSPs of a metallic nanoparticle (NP) - energy, field enhancement and field distribution - are determined by its geometry, dielectric properties and, crucially, its proximity to other (metallic) structures, the issue specifically addressed in this work.

In the present study we focus on junction based optical spectroscopy techniques like light emission from scanning tunnelling microscopy (LESTM) and TERS. Both exploit the high electric field enhancement associated with the LSPs of a 'nanocavity' (formed between a metallic tip in close proximity of a substrate) and have been used to probe the energetics of nanoentities placed within, from molecules to nanowires. Enhanced sensitivity, originating with the strong field enhancement, is combined with high spatial resolution that stems from the localized nature of the LSPs in the nanocavity. TERS has shown chemical identification of single molecules [8] and nanowires [9] with a lateral resolution (LR) of 10–20 nm, while chemical sensing in LESTM has been performed with a resolution of a few nanometers [10]. The geometry of these experiments is treated here as that of a metal NP (representing the tip) in close proximity to a substrate (figure 1(a)). A number of theoretical studies have investigated the LSP modes of both the tip–sample [11,[12,[13] and the NP–substrate system [14,[15,[16]; however they have tended to be either pure modeling exercises (i.e. no direct comparison to experiment is drawn) and/or do not mimic the specific conditions of LESTM. In this investigation we extend the analysis of such studies by modeling the exact conditions of the LESTM set-up (i.e. tip and sample in *tunnelling* proximity) and by producing detailed correlation of the modeled data with experiment.

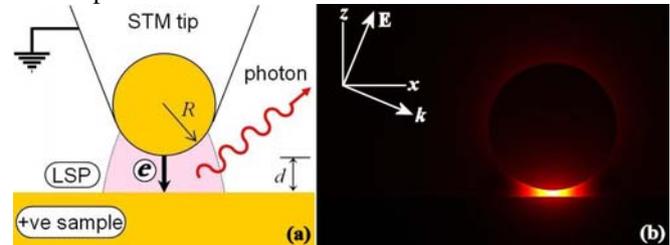

Figure 1: (a) Schematic of the STM tip-sample junction (b) *xz* plane snapshot of the calculated electric field distribution around a 20 nm diameter Au nanoparticle near a 30 nm thick Au substrate for $\lambda$ = 900 nm. The propagation vector *k* is 20º to the *x*-axis.

We have applied finite element method analysis (FEMA) to identify the LSP modes of a single spherical Au NP (of radius *R*), within tunnelling proximity (gap dimension *d*) of a semi-infinite planar Au substrate. Using plane wave optical excitation, individual LSP modes are evidenced as peaks in plots of the electric field enhancement factor ($E_f$) vs. wavelength ($\lambda$) for the nanocavity. Using a simple reciprocity argument, LSP modes excited in the nanocavity should couple to waves propagating away from the nanocavity region, with spectra of the outcoupled light exhibiting peaks at energies corresponding to those of the resonant LSPs. Here we make the connection to LESTM (figure 1(a)), in which tunnel current fluctuations excite the LSP modes of the tip – sample nanocavity [17,[18,[19], and subsequently decay, giving rise to broadband emission with characteristic peaks at the modal energies; optically, the NP in the modeling equates to the scanning tunnelling microscope (STM) tip and should be of closely comparable radius. Though the excitation mechanism is different in the two scenarios the energies of the LSP modes (being purely a function of the nanocavity geometry and dielectric properties of the environment) remains unchanged.

In the following we demonstrate a close match between the LSP modal energies obtained in the experimental and simulated results. While confirming a large red-shift of the LSP modal energies with increasing *R* the results also elucidate the spatial localization of $E_f$ in the nanocavity. We consider the limit of the LR achievable in LESTM and the conditions required for its implementation. These findings are highly relevant to the design of any junction based optical spectroscopy.

## 2. Material and Methods.

The wavelength-dependent response of the NP–substrate system was simulated using a commercial 3D FEMA package (COMSOL Multiphysics 3.3a). The simulation domain comprises a single spherical NP separated from a semi-infinite square plane of thickness 30 nm by a gap with $d = 0.6$ nm (a typical value for STM). A thickness larger than the skin depth of Au is chosen to exclude any thickness dependent effects of the LSP excitation energies [14]. To maintain the semi infinite nature of the plane, the ratio of NP radius to plane dimension was fixed at 1:10. The Au NP and substrate is characterized by Drude dielectric function given by $\varepsilon_{Au} = \{1 - \omega_p^2/(\omega^2 + \gamma^2)\} + i\{\gamma\omega_p/(\omega^2 + \gamma^2)\}$, where $\hbar\omega_p$ is the bulk plasmon frequency and $\gamma$ the damping term. The surrounding medium is considered as free space ($\varepsilon = 1.00$). The model is non-uniformly meshed ensuring approximately 10 mesh points between the NP and substrate. The simulation calculates the local electromagnetic field at every mesh point with an incident p-polarized plane wave (figure 1(b), propagation vector tilted at 20° below the x-axis) for the wavelength range 400 - 1500 nm at 10 - 20 nm intervals. The incident electric field strength is scaled to unity so that the calculated field represents $E_f$. The values used for $\hbar\omega_p$ and $\gamma$ are 3.6 eV and 0.1 eV respectively; we comment further on the value of $\hbar\omega_p$ in the discussion of the results.

The experimental set-up comprises a Digital Instruments Nanoscope-E STM operated in ambient with Au tips on Au(111) surfaces. We present emission spectra corresponding to three Au tips (prepared by electrochemical etching of Au wire [20]) with tip end diameters of ~ 20, 50 and 160 nm, as estimated from scanning electron microscope images. Au(111) surfaces were freshly prepared by flame annealing 200 nm thick Au films. The light output was collected by two 800 μm-core, low-OH⁻ content optical fibers, positioned ~1 mm from the tip-sample region and fed to an Acton Research SpectraPro 275 spectrometer fitted with an Andor DU420-OE charge coupled device (CCD) camera to record the emission spectrum (overall detection range 400 – 1000 nm). In this setup the long wavelength cutoff, of the recorded spectra, is dictated by the CCD camera ($\lambda_{max}$ ~ 1000 nm), while the short wavelength cutoff, $\lambda_{min}$, is determined by the applied sample bias ($V_b$) according to the relation $hc/\lambda_{min} = eV_b$. The spectral data were acquired for a positive sample bias of ~ 1.8 V ($\lambda_{min} \approx 688$ nm) and 10 nA tunnel current. Each spectrum was obtained by averaging ten 60 s exposures. The spectra are not corrected for the wavelength dependent efficiency of the detection setup; such correction changes the peak positions by less than 10 nm (illustrated in figure 3, where the shift is actually < 5 nm).

## 3. Results and Discussion.

Figures 2–4 plot the LESTM spectra from tips with end radii, $R = 10$, 25 and 80 nm with the corresponding FEMA simulation data, showing plots of the normalized $E_f$, at a point (0.1 nm below) directly below the NP. The spectrum in figure 2 (for tip with $R \sim 10$ nm) shows a single peak at $\lambda \sim 900$ nm, which is directly reflected in the $E_f$ vs. $\lambda$ plot with a peak at 895 nm. The peak at $\lambda \sim 680$ nm in $E_f$ is not observed experimentally due to the short wavelength cutoff imposed by the sample bias. It is worth noting that for this sharp tip the simulation indicates the absence of any further peaks in $E_f$ at longer wavelengths [19], i.e. the lowest energy LSP mode occurs at $\hbar\omega_0 = 1.39$ eV (895 nm).

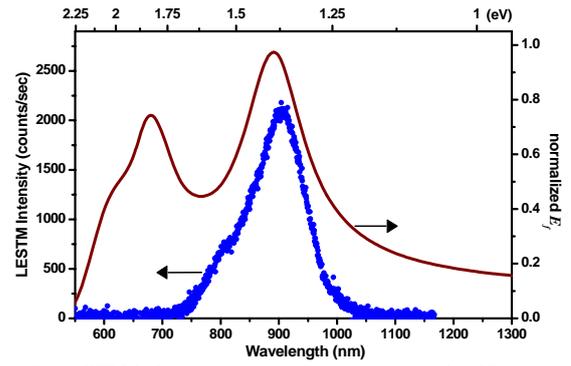

Figure 2: (•) STM light emission spectrum for tip of $R \sim 10$ nm for $V_b = 1.75$ V ($\lambda_{min} \sim 709$ nm) and corresponding simulation showing normalized $E_f$ as a function of wavelength (energy).

Figure 3 (for $R \sim 25$ nm tip) also shows a single-peaked ($\lambda \sim 770$ nm) spectrum that is replicated in the corresponding $E_f$ data which exhibits a peak at 765 nm; corresponding to the first higher order mode with $\hbar\omega_1 = 1.62$ eV. The $E_f$ plot predicts that the lowest energy LSP mode has now shifted beyond our detection range to $\hbar\omega_0 = 1.17$ eV (1059 nm). Figure 4 shows the STM emission spectrum from the third tip ($R \sim 80$ nm) with two distinct peaks at 774 nm and 893 nm with corresponding maxima in the $E_f$ plot at 779 nm and 891 nm. In this case the simulation does not extend to wavelengths that cover the lowest energy LSP that occurs for $\lambda > 1200$ nm.

The surface plasmon resonance wavelength of an isolated NP (~ 550 nm for Au NPs [21]) is known to red-shift when placed in close proximity to a metal substrate, the red-shift increasing with increasing proximity and with NP radius [22]. The larger red-shifts observed here, even for the smallest NP with the spectral peak at 900 nm is a combined effect of the NP–substrate proximity ($d = 0.6$ nm) and a *local* reduction in $\hbar\omega_p$ for Au (3.6 eV as used in the simulations).

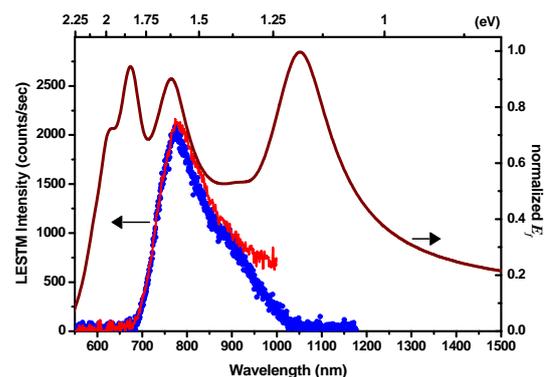

Figure 3: (•) STM light spectrum for tip of $R \sim 25$ nm and (-) spectrum after correction for the efficiency of the spectrometer-detector setup for $V_b = 1.80$ V ($\lambda_{min} \sim 688$ nm). Corresponding simulation showing normalized $E_f$ as a function of wavelength (energy).

Within the Drude model this reduced value of $\hbar\omega_p$ can be interpreted as an effective lowering of the average density of conduction electrons. For two metal bodies within tunneling proximity, with large background polarizability and electric fields involved would lead to surface charge screening, constraining the number of free electrons available to participate in LSP oscillations [19]. This effect would be restricted to the immediate vicinity of the nanocavity. On the other hand, indications are that for $\lambda > 600$ nm the intrinsic optical properties of *isolated* Au NPs show limited deviation from bulk values [23].

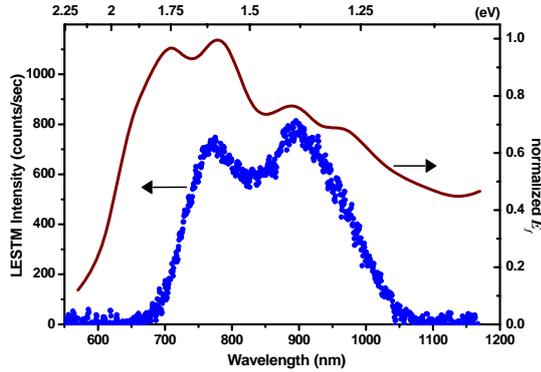

Figure 4: (●) STM light emission spectrum for tip of $R \sim 80$ nm for $V_b = 1.80$ V ($\lambda_{min} \sim 688$ nm), and corresponding simulation showing $E_f$ as a function of wavelength (energy).

A comparison of figures 2-4 shows that with increasing NP radius the LSP peaks shift progressively further into the infrared. Also, the overall magnitude of $E_f$ increases monotonically with $R$ (the maximum $E_f$ increases four fold between $R = 10$ and $80$ nm) and assumes a more broadband nature (for the range explored here). These features are highly pertinent to the design of any junction-based optical spectroscopy. To exploit the LSP mediated signal enhancement a region of significant field enhancement must span the relevant energy range of interaction (e.g. the vibrational energy or the Stokes line energy of any molecule in the nanocavity). Thus, the broader the line-width of the LSP modes or the higher the uniformity (across $\lambda$) of the overall enhancement of a junction, the greater is its applicability to a varied range of molecules.

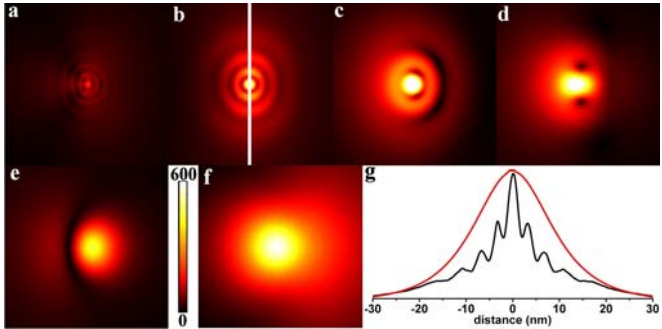

Figure 5: (a-e) 30×30 nm plot of $E_f$ on a $xy$ plane, centered 0.1 nm below an Au nanoparticle ($R = 80$ nm), with an Au substrate, for $\lambda = 600, 700, 800, 900$ and $1200$ nm respectively. The same colour scale shown is used for (a-e). (f) – same as (a), but with SiO$_2$ substrate (colour scale is × 15 that of a-e). (g) – plot of normalized $E_f$ along the line $x = 0$ (depicted in b) for image a (black) and e (red).

A crucial aspect of both TERS and LESTM spectroscopy is the LR afforded by the techniques. This is determined by the spatial variation of the electric field within the nanocavity, specifically in the $xy$ plane. For the case of the R = 80 nm NP, figure 5(a-e) shows a series of 2D plots of $E_f$ across a 30 nm square $xy$ plane, positioned 0.1 nm directly below the NP. The snapshots at five specific $\lambda$ (600, 700, 800, 900 and 1200 nm) clearly show an extremely non-uniform oscillatory electric field in the nanocavity (more distinct for lower values of $\lambda$) progressively decaying with increased distance from the system axis.

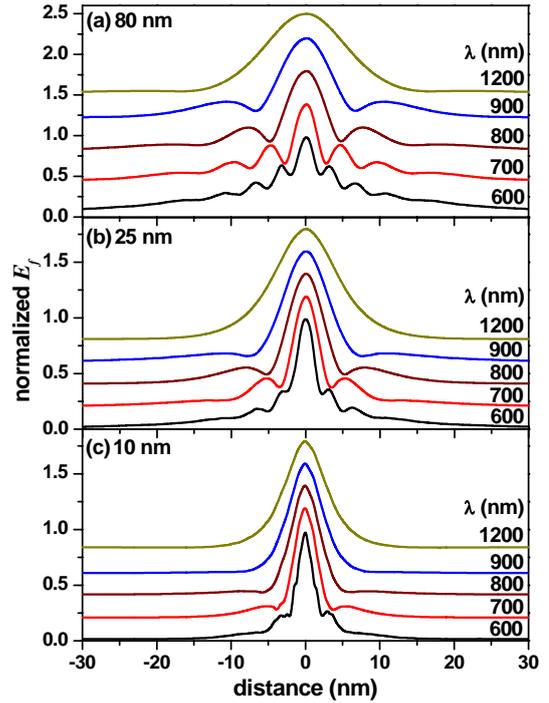

Figure 6: Normalized $E_f$ plotted along the line shown in figure 5(b) for NP radii (a) 80 nm, (b) 25 nm and (c) 10 nm for various values of $\lambda$. Individual plots are vertically shifted for clarity.

The asymmetric nature of $E_f$ (figure 5) along the $x$-direction is a consequence of the asymmetry in the incidence of the excitation field. Figure 6(a)-(c) shows plots of normalized $E_f$ (along the line shown in figure 5(b)) at specific wavelengths (see above) for NPs of radii 80, 25 and 10 nm. The oscillatory variation of $E_f$ with distance from the system axis is a feature specifically associated with the *proximity* of the NP to a *metallic* substrate. This is clearly demonstrated by comparison with the case where the substrate is replaced with a *dielectric* medium, SiO$_2$, as shown in figure 5(f); this is simulated for the same parameters as for 5(a) but with the substrate having a refractive index of 1.45. The contrast is further elucidated in figure 5(g); that plots the normalized $E_f$ from figures 5(a) and (f) along the line $x = 0$. For the case of SiO$_2$, the lateral variation assumes a broader, typically Gaussian profile [7,[13]. At $\lambda = 600$ nm the magnitude of $E_f$ for the Au substrate is ~ 7 fold stronger than for the SiO$_2$ substrate. The stronger enhancement and the oscillatory variation of $E_f$ observed for an Au substrate originates with a stronger resonant coupling between the NP LSPs and the propagating plasmons of the substrate surface.

Figure 6 shows that the central peak width is a function of both $R$ and $\lambda$. However the presence of the secondary peaks (and their decay length scale) significantly affects any calculation of the LR (discussed below). It is worth noting that for a specific NP the peak to peak separation ($s_{pp}$), is found to increase monotonically with $\lambda$. $s_{pp}$ varies from 3 nm at $\lambda = 600$ nm to 22 nm at 1200 nm for the 80 nm NP, while for the 10 nm NP $s_{pp}$ varies from 1.5 nm at 600 nm to > 10 nm at 900 nm. The values are highly relevant to NP assisted imaging and lithography applications.

Figure 7 plots the normalized $E_f$ value at the peak positions for the various excitation wavelengths $600 \leq \lambda \leq 1200$ nm for all three NPs and is used to determine the LR. The peak $E_f$ values follow a first order exponential decay (dashed curves in figure 7). The decay length is primarily governed by the geometric properties of the nanocavity (parameters $R$ and $d$) and, as expected, is least for the smallest NP. Interestingly, there is negligible wavelength dependence within the range probed. For TERS the LR is determined by the variation of the fourth power of $E_f$ and has been discussed extensively in literature [11,[13]. Here, in the context of LESTM, we calculate the LR as the distance at which $|E_f|^2$ falls to 50% of its central maximum. The variation of this calculated LR with NP radius is shown in the inset to figure 7. A fit to the LR values closely follows a $R^{1/2}$ dependence that mimics the LSP localization length dependence on $R$ predicted earlier [15]. However, even for the $R = 10$ nm NP, the LR ~ 1.2 nm and the simulations predict sub-nanometer resolution only with tips of $R < 5$ nm.

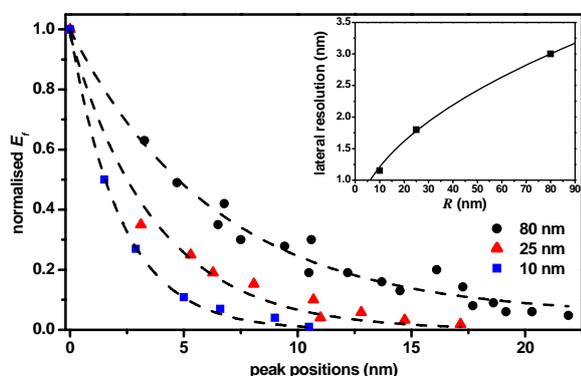

Figure 7: Dependence of the normalized peak electric field values with peak positions for nanoparticles of radius $R = 10$, 25 and 80 nm; displaying first order exponential decay (dashed lines). Inset shows the calculated lateral resolution as a function of nanoparticle radius and the fitted line indicates a $R^{1/2}$ dependence (see text).

The above analysis of achievable LR pertains to LSPs and the local variation in the electromagnetic coupling. Variation in the electronic structure may occur on a more localized scale and be a sufficiently dominant influence to yield higher resolution optical contrast. Thus Hoffman et al. [24] explain the contrast in atomically resolved photon maps (on Au(111) surface) purely in terms of local variation in electron tunnelling probability, neglecting any lateral variation in electromagnetic coupling.

## 4. Conclusions

In conclusion we have performed finite element calculations of the electric field enhancement in an Au nanoparticle–substrate nanocavity as a function of wavelength and nanoparticle radii. The calculated LSP modes of the nanocavity can be experimentally excited and analyzed by recording the LESTM spectra from Au tips of comparable radii. Agreement between the calculated LSP modal energies and those obtained experimentally dictates a reduced effective bulk plasma frequency for Au. The simulation results also show that with increasing nanoparticle radii the LSP modes red shift and the overall electric field enhancement increases significantly in magnitude with an almost broadband enhancement shown for the largest nanoparticle ($R = 80$ nm). In contrast to earlier reports the lateral variation of the electric field in the nanocavity is found to be oscillatory in nature with an exponential lateral decay. A consequence of the tunnelling proximity of the nanoparticle–substrate system and the metallic nature of the substrate. The spatial resolution is shown to be primarily governed by the geometric properties of the nanocavity and is limited to supra-nm scale even for the smallest nanoparticle investigated. Sub nanometer resolution is predicted for still sharper tips of end radius < 5 nm.

## Acknowledgements
We thank Mr. Anthony Murphy for helpful discussions regarding COMSOL. The authors acknowledge financial support from the UK Engineering and Physical Sciences Research Council and 'Nanotec Northern Ireland' supported by EC funding through Invest Northern Ireland.